\PassOptionsToPackage{table}{xcolor}
\PassOptionsToPackage{hyphens}{url}
\documentclass[letterpaper,twocolumn,10pt]{article}
\usepackage{usenix-2020-09}

\usepackage{tikz}
\usepackage{amsmath}
\usepackage[utf8]{inputenc}
\usepackage{hyperref}
\usepackage{url}
\usepackage{todonotes}
\usepackage{cite}
\usepackage{comment}

\usepackage{xspace}
\usepackage{wasysym}
\usepackage{amssymb}
\usepackage{adjustbox}
\usepackage{titlesec}
\usepackage{enumitem}
\usepackage{tcolorbox,xstring,xspace,etoolbox} 
\usepackage[table]{xcolor}
\usepackage{colortbl}
\usepackage{sidecap}
\usepackage{pdfpages}

\usepackage[available]{usenixbadges}

\usepackage{array}%
\newcolumntype{R}[1]{>{\RaggedLeft\arraybackslash}p{#1}}
\newcolumntype{X}[1]{>{\centering\arraybackslash}p{#1}}
\newcommand{\rot}[1]{\multicolumn{1}{c}{\adjustbox{angle=60,lap=\width-1em}{#1}}}

\usepackage{caption}
\usepackage{subcaption}
\usepackage{censor}  %

\ifdefined\BLINDED
	\def\toolname{\textsc{[Redacted]}\xspace}
	
	\makeatletter
	\def\gobblechar{\let\xchar= }
	\def\assignthencheck{\afterassignment\xloop\gobblechar}
	\def\xloop{%
	  \ifx\relax\xchar
      	\let\next=\relax
	  \else
    	\ifx\@sptoken\xchar%
	    	\xchar\let\next=\assignthencheck
	    \else
	    	x\let\next=\assignthencheck
	    \fi
	  \fi
	  \next}
	\def\xxx#1{\assignthencheck#1\relax}

	\def\ifBLINDED{\expandafter\@firstoftwo}
	\makeatother
	
\else

	\def\toolname{\textsc{MobileAtlas}\xspace} %
	\StopCensoring %
	\def\xxx#1{#1}
	
	\makeatletter
	\def\ifBLINDED{\expandafter\@secondoftwo}
	\makeatother
\fi

\xspaceaddexceptions{.}\xspaceaddexceptions{,}
\newcommand{\roundframe}[1]{{\setlength\fboxrule{0pt}\fbox{\tcbox[colframe=black,colback=white,shrink tight,boxrule=0.5pt,extrude by=2.5pt]{\small #1}}}}

\begin{document}

\title{\Large \bf \toolname: Geographically Decoupled Measurements \\ in Cellular Networks for Security and Privacy Research}

\date{}

\author{
{\rm \xxx{Gabriel K. Gegenhuber}}\\
\xxx{University of Vienna}\thanks{Supported by the UniVie Doctoral School Computer Science DoCS.}
\and
{\rm \xxx{Wilfried Mayer}}\\
\xxx{SBA Research}
\and
{\rm \xxx{Edgar Weippl}}\\
\xxx{University of Vienna}
\and
\and
{\rm \xxx{Adrian Dabrowski}}\\
\xxx{CISPA Helmholtz Center for Information Security}\thanks{Partly as postdoc at University of California, Irvine.}
} %

\emergencystretch3em
\setlength{\parskip}{0pt plus 3pt}
\setlength{\itemsep}{0pt plus 1pt}

\setlength{\belowcaptionskip}{7pt}
\setlength{\abovecaptionskip}{3pt}

\setlength{\floatsep}{6pt plus 5pt}
\setlength{\textfloatsep}{10pt plus 10pt}

\widowpenalty1
\clubpenalty1

\pagenumbering{gobble} %

\titlespacing*{\subsubsection}{0pt}{1ex plus 5pt}{0.5ex plus 2pt minus 2pt}
\titlespacing*{\subsection}{0pt}{2ex plus 1ex}{0.9ex plus 3pt minus 3pt}
\titlespacing*{\section}{0pt}{1.5em plus 1em}{0.8em plus 1ex minus 1ex}

\makeatletter
\def\paragraph{\@startsection{paragraph}{4}{1\parindent}{0pt plus 3pt}%
{-1\parindent}{\normalfont\normalsize\bfseries}}%
\makeatother

\def \ifempty#1{\def\temp{#1} \ifx\temp\empty }
\newcommand{\TODO}[1][]{\colorbox{yellow!80}{\ifempty{#1}\textbf{TODO}\else\textbf{TODO}: #1\fi\xspace}}

\maketitle

\begin{abstract}%
    Cellular networks are not merely data access networks to the Internet. Their distinct services and ability to form large complex compounds for roaming purposes make them an attractive research target in their own right. Their promise of providing a consistent service with comparable privacy and security across roaming partners falls apart at close inspection. 

Thus, there is a need for controlled testbeds and measurement tools for cellular access networks doing justice to the technology's unique structure and global scope. Particularly, such measurements suffer from a combinatorial explosion of operators, mobile plans, and services.
To cope with these challenges, we built a framework that geographically decouples the SIM from the cellular modem by selectively connecting both remotely.
This allows testing any subscriber with any operator at any modem location within minutes without moving parts. The resulting GSM/UMTS/LTE measurement and testbed platform offers a controlled experimentation environment, which is scalable and cost-effective. The platform is extensible and fully open-sourced, allowing other researchers to contribute locations, SIM cards, and measurement scripts.

Using the above framework, our international experiments in commercial networks revealed exploitable inconsistencies in traffic metering, leading to multiple \textit{phreaking} opportunities, i.e., fare-dodging. 
We also expose problematic IPv6 firewall configurations, hidden SIM card communication to the home network, and fingerprint dial progress tones to track victims across different roaming networks and countries with voice calls.

\end{abstract}

\begin{figure*}[!t]%
	\centering%
	\includegraphics[width=0.95\linewidth]{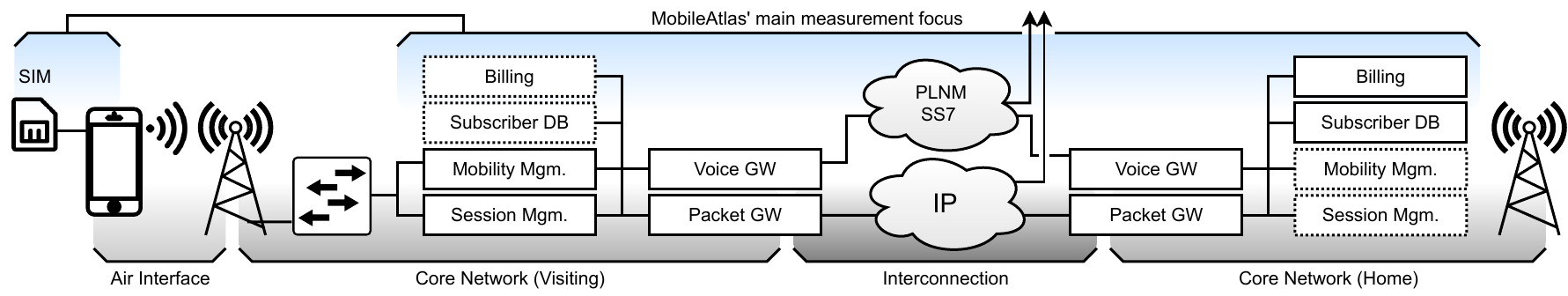}%
	\caption{Simplified technology neutral structure of a cellular 3GPP network with roaming, and \toolname's primary focus}%
	\vspace{-5mm}%
	\label{fig:mnostruct}%
\end{figure*}

\section{Introduction}\label{sec:introduction}

Large-scale measurement platforms and distributed testbeds such as RIPE Atlas \cite{staff2015ripe} and PlanetLab \cite{planetlab} contributed to the security and privacy research community in at least three ways: 
(i) they allow the measurement of network properties once or longitudinal from different vantage points,
(ii) they allow to quickly measure the scale of a found or known problem, i.e., gauge the real-world impact, and
(iii) they function as a testbed to rapidly develop and test potential security vulnerabilities on a large scale.
Additionally, tools such as ZMAP \cite{Durumeric2013zmap} provide the ability to routinely make Internet-wide scans, which became a staple for papers on measurement and security alike.

These platforms and tools share that --in accordance with the layered network model-- they are access technology agnostic. However, mobile networks, unlike any other access network, combine multiple access technologies and generations on top of each other. 
Furthermore, since Mobile Network Operators (MNOs) are only given a small geographical area (usually a country) to operate in, they form vast roaming alliances to allow devices (and their traffic) to traverse through multiple networks. This creates complex compound systems where entities of different operators handle different aspects of the user traffic.

To explore such systems, physically moving devices (or SIM cards) between countries for each case adds a staggering, prohibitive overhead. 
MONROE \cite{monroe-project} approached this problem by duplicating each SIM card set at each location -- effectively realizing the combinatorial explosion of \mbox{\textit{countries} $\times$ \textit{mobile plans}} \textit{for each operator} -- with tremendous costs hindering growth.

In this paper, we present a different approach. The key insight is that by geographically decoupling the SIM from the device, we can work with just one set of devices in the field and virtually connect them to one set of SIM cards -- without multiplying our SIM arsenal for each location.
To this end, we tunnel the SIM card protocol from one of the SIM cards to any of our remote-controlled devices in different countries and territories and have them connect selectively to specific operators as if the subscriber SIMs were actually there. 

This enables us to test security and privacy-relevant aspects of different operators and combinations of operators as well as perform large-scale international studies and investigations.

We implemented the method in \toolname, based on low-cost off-the-shelf hardware to facilitate easy deployment, deployed it in ten countries, and added an orchestration \& management interface to encourage other researchers to join.
As an open and scalable tool, we hope it will help researchers, but it could also be used by operators or regulatory bodies to validate local or roaming network properties. \toolname is versatile enough to handle data, voice, SMS, and USSD\footnote{Unstructured Supplementary Service Data codes are user dialable messages to control provider services, e.g., \#\#21\# to disable all call forwarding.} services as well as capture hidden SIM communication.%

We demonstrate the capabilities and flexibility of this framework with five security- and privacy-related use cases in three different domains: (i) high-scalable, isolated network measurements in domestic \& roaming scenarios, 
(ii) voice connections with and between roaming partners, 
and (iii) low-level SIM communication analysis. The results demonstrate how to leak a traveling callee's country and operator as well as transmit free data by bypassing traffic accounting.

The remainder of this paper is structured as follows: Section~\ref{sec:background} and \ref{sec:relatedwork} introduces the background and related work.
In Section~\ref{sec:analysis}, we analyze the problems and requirements of measurement systems.
In Section~\ref{sec:system-design}, we describe the general system design, followed by the concrete implementation in Section~\ref{sec:implementation}.
We present current use cases in Section~\ref{sec:showcases}, %
discuss our work in Section~\ref{sec:discussion}, and conclude in Section~\ref{sec:conclusion}.

\section{Background}\label{sec:background}

\subsection{Cellular Networks and Roaming}
\label{sec:background_cellularnetworks}

For the sake of space, we describe a 3GPP-style cellular network (i.e., GSM, UMTS, LTE, 5G) in a simplified structure and with technology neutral terminology (Figure \ref{fig:mnostruct}). 

The \textit{SIM card} (physical or as embedded SIM) is issued by the home operator and holds secret cryptographic keys. 
Any device (i.e., UE, User Equipment) successfully passing authentication based on those keys is recognized as a specific subscriber. The SIM itself does not encrypt the data over the air interface (RAN, radio access network), it only provides the session keys. 

An MNO operates a large number of base stations\footnote{\textit{Base Transceiver Station BTS} in GSM, \textit{NodeB} in UMTS, \textit{E-UTRAN Node B} or short \textit{eNodeB} in LTE, \textit{gNodeB} in 5G: The reader may excuse that we will use a high-level technology neutral terminology to focus on the structure and not on a particular implementation.} that are  connected via the core network to centralized or decentralized services. 
A subscriber database and a billing system manage the customers at the home operator. A session management system allocates network services of a given quality for a given subscriber, e.g., Internet access at a set speed over a high-latency tunnel, a voice channel on a low-latency low-bandwith tunnel/channel, a phone number, etc. 
A mobility management unit keeps track of the UE's position within the network and can page (call out) the UE if necessary. It makes sure the session tunnels are rerouted accordingly when the UE changes location. 

\subsubsection{Services}
Today, cellular networks offer three main services to customers: data, voice, and SMS (i.e., text). 

Data services are provided via a data tunnel (\textit{bearer} since LTE) that follows the subscriber and (typically)  terminates the traffic via a public-Internet-facing gateway. 

In GSM and UMTS, voice is a different network service than data. It uses a different type of tunnels and channels (\textit{circuit-switched}), priorities, and endpoints. 
On LTE and 5G, voice (Voice over LTE, VoLTE) is IP traffic and transported via the same type of tunnels as Internet data, although with a different priority. Those tunnels terminate at an IMS-Server (IP Multimedia Subsystem) that can route calls to the public phone network. 
Networks often operate two separate voice infrastructures, one for VoLTE and one for legacy circuit-switched calls. 
The network can use Circuit-Switched Fall Back (CSFB) to move 5G/LTE voice calls to UMTS or GSM.

\subsubsection{Roaming} 
\label{sec:background_roaming}
Frequencies for cellular networks are typically given exclusively to a selected number of operators in each country or territory. To provide services outside of the physical coverage area, an MNO arranges a roaming agreement with one or more MNOs in another area. 

In the roaming case, a UE uses a mixture of services from the two involved operators: the \textit{visiting operator} (where the subscriber is physically located) and the \textit{home operator} (where the subscriber got their SIM from). From the subscriber's view, both operators fuse to a single system that strives to provide the same services as the home operator in their home territory. 

\paragraph{Data}
Internet data can be roamed in two ways: (i) \textit{Local Breakout} (LBO) terminates the traffic at the visiting operator's Internet gateway. A customer will therefore receive an IP address from their visiting country which may effect website localization and geoblocked services. Billing records are collected by the visiting operator.
(ii) With \textit{Home Routing} (HR), the visiting operator hands over all subscriber data traffic to the home operator (usually via an Internet interconnection), where it is billed and routed to the Internet with a home operator's IP address.

\paragraph{Legacy Voice and LTE's CSFB}
The visiting operator assigns a temporary local phone number, i.e., from the visited country. Voice calls take the same infrastructure as the visiting operator's own subscribers, with the exception that the outgoing caller ID is matched to the original visiting customer's number. In this regard, it is similar to the local breakout data roaming case.
Incoming calls are handled by the home operator, which redirects the call to the temporary foreign phone number by opening another phone call. 

\paragraph{VoLTE roaming}
VoLTE roaming can be handled as LBO, HR, or a combination of those that involves both IMS. 
However, because the complexity of the underlying SIP protocol and the large number of settings, e.g., for bitrate and codec, the standardization and interoperability is still in its infancy. 
Many operators use \textit{Circuit-Switched Fall Back} (CSFB) for roaming, and some (such as AT\&T in the U.S.) stopped providing voice roaming altogether.

\subsection{Types of Measurements}
\label{sec:background_measurement_types}

For measurements in the mobile network sector, we differentiate between (i) passive and (ii) active measurements.
Passive measurements often require elevated network access and are therefore collected with the help of network operators, e.g., Mobile Network Operators (MNOs) or Internet Service Providers (ISPs). This data is seldom publicly available. 

In contrast, many active measurements work with a customer-grade access and are a vital path for independent and open research.
The two most common practices for active measurements include (i) in-situ or in-vivo measurements and (ii) exclusive measurement setups.

\textit{In-situ-} (``at location'') and \textit{in-vivo} (``within the system'') measurements often run on a (volunteer's) phone that is also used for other tasks.
This method increases coverage by lowering the financial burden of participating in a given study and thus adds more measurement units.
While appropriate for, e.g., user studies, this method might impact the accuracy of technical measurements since the non-exclusivity blurs the distinction between the signal (that is to be measured) and the noise (created by user activity or background processes).
The user's mobility also interferes with the common goal of steady (consistent and repeatable) measurement environments. 

Exclusive measurement setups require a separate test unit in a controlled environment.
While some external factors (such as the network's utilization or radio noise) are commonly outside the researchers' sphere of influence, other factors can be controlled; this includes the distance to a cell tower, the direction of the antenna, and background data usage.

\section{Related Work}\label{sec:relatedwork}

\paragraph{Measurements for Security}
Measurement of the impact of a found vulnerability is part of many network-related security papers. 
ZMAP \cite{Durumeric2013zmap} and other similar tools became a staple for evaluation of Internet phenomena and especially security vulnerabilities.

\paragraph{Measurement of and in cellular networks}

An example for a passive measurement is Lutu et al.'s~\cite{lutu2021insights} insights into a large IP exchange provider (IPX-P).
They analyzed passively collected data and gathered operational insights regarding, among others, global data roaming.

In 2019, Li et al.~\cite{li2019large} %
developed an active \textit{in-vivo} mobile application called \emph{Wehe} to gather more than one~million crowdsourced measurements.
They identified traffic differentiation at 30~ISPs.
\toolname can, in addition, use further mobile network capabilities to conduct measurements, e.g., SMS, calls, and USSD. It provides a controlled measurement environment with no background noise from other apps.
Thus, our technique is well suited for longitudinal observations under stable conditions and can measure fine-grained differentiations. %

\begin{figure}[!t]%
    \centering
    \includegraphics[width=\linewidth]{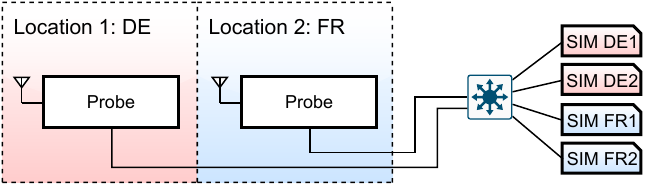}
    \caption{Decoupling the station from the SIM via tunneling requires only one set of SIMs.}   
    \label{fig:sharing-approach}
\end{figure}

In 2015, Alay et al.~\cite{alay2015monroe} presented the MONROE measurement platform and later refined the architecture and node design~\cite{alay2016measuring}. 
MONROE is an open, European-scale, and flexible measurement platform for mobile broadband networks.
However, each node is physically connected to the measured SIM card, leading to maintenance and hardware overhead.
Similar to \toolname, MONROE is an exclusive measurement platform that also supports roaming. 
However, \toolname improves scalability by reusing one single SIM card at different nodes.
We contrast those architectures in Figures~\ref{fig:traditional-approach} and \ref{fig:sharing-approach}. %

\paragraph{Distributed Measurement Networks}
Landline and fixed Internet measurement platforms matured over the decades and are a staple of empirical Internet research. 
For instance, the RIPE Network Coordination Centre introduced the RIPE Atlas measurement platform~\cite{ripe-atlas, staff2015ripe} in 2010, where a large number of simple measurement probes allows a distributed execution of a small set of network commands, e.g., traceroute.
Users can create self-defined measurements that get accounted for through platform-specific credits.
RIPE Atlas has continuously improved~\cite{bajpai2015lessons} over the last decade and consists of approximately 12,000 measurement \emph{probes} and 800 \emph{anchors} (February 2023). %
It is utilized by numerous research projects, covering issues such as 
network routing information~\cite{anwar2015investigating}, various protocol measurements~\cite{jones2016detecting} and censorship detection~\cite{anderson2014global}.
However, RIPE Atlas is focused on fixed Internet connections and only allows a small set of possible measurement commands -- \toolname replicates these (and other) capabilities in the cellular world.

Kakhki et al.~\cite{kakhki2015identifying, kakhki2016bingeon} studied traffic differentiation in mobile networks, using an app-based record-replay approach to identify traffic differentiation based on mobile application traffic.
\toolname could improve these results because our probes are fully controllable and the provisioning of different SIM cards to test geographical differentiation is possible.

\paragraph{Voice Interconnection Security and Fraud}
In 2016, Sahin and Francillon~\cite{sahin2016over,Sahin2017} measured so-called \textit{over-the-top} (OTT) bypass fraud, i.e., telephone calls that get fraudulently redirected to OTT services, bypassing terminating operators.
The authors detected OTT fraud using Test Call Generation (TCG) networks to initiate tens of thousands of phone calls.
This approach is helpful for incoming bypasses, whereas outgoing routes would require an international network.
In addition to their TCGs, Sahin et al. had to geographically move phones with their SIM cards between different countries and operators. %
\toolname supports voice calls and can therefore automate OTT and other fraud~\cite{sahin2021understanding} detection.

\paragraph{Decoupling and Redirecting cellular subscribers}
Samsung Call \& Message Continuity (CMC) \cite{samsung-cmc} is a way to seemingly transfer calls and chats between devices on the fly. However, CMC does not change the transfer point with the cellular network, it just acts as a relay to forward data. 
SIM Banks or SIM Pools are mainly used in OTT fraud \cite{sahin2016over} to distribute and shuffle subscriber SIMs between different exit points domestically via SIM tunneling to circumvent roaming. 
In contrast, \toolname moves SIMs internationally, to actively trigger roaming. SIM Banks are occasionally available from AliExpress and other sites but use an undocumented transmission protocol. 

\paragraph{Air Interface Attacks}
Cellular network research is often narrowly considered as research of the air interface or radio access technology. Certainly, \textit{Fake Base Stations} (FBS) or \textit{IMSI Catchers} \cite{Dabrowski2014icc, vanDo2015, Ney2017 , Park2017, Alrashede2019IMSICatcher} are one of the most prominent ones. FBS are a major building block for a large number of further cellular network exploits \cite{RupprechtDabrowski2018cst}. 
\toolname can partly work as Fake Base Station detection, but would need a much denser distribution pattern to cover single base stations (Section \ref{sec:deployment}) and is therefore not a goal at this stage (Figure~\ref{fig:mnostruct}). \looseness=-1

\section{Analysis}\label{sec:analysis}

\begin{figure}[!t]%
    \centering
    \includegraphics[width=\linewidth]{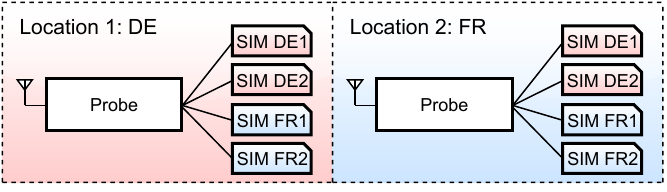}
    \caption{Traditional approach with poor scalability: Every new location needs a new set of all SIMs and mobile plans.}
    \label{fig:traditional-approach}
\end{figure}

As briefly touched above, current measurement platforms suffer from one or more of the following problems or challenges:
\begin{itemize}[topsep=3pt plus 2pt,itemsep=0ex]\setlength{\itemsep}{3pt plus 2pt}\setlength{\parskip}{1pt plus 2pt minus 2pt}%
	\item[\roundframe{P1}] \textit{Access technology obliviousness.}~  
	Some measurement systems are ignorant regarding the underlying properties and features of the access network.
    In many cases, the transparent layering of a network stack from the physical layer up to the application layer is a desirable property, e.g., receiving e-mail should work regardless of Wi-Fi, 3G, LTE, or roaming.
    However, for measurements, the layering may hide crucial information and defining characteristics.
   
   Without full access to domain-specific parameters (e.g., Cell- \& Location IDs) and features (e.g., voice calls), a measurement system may lack sufficient insights to successfully audit complex mechanisms (e.g., roamed billing) within cellular networks.

\item[\roundframe{P2}] \textit{Resources scale quadratically by coverage.}~ %
  If a cellular device and a mobile plan (mediated through the SIM card) are treated as a single unit, they are location-bound.
  Devices and mobile plans (i.e., SIM cards) need to be duplicated at every location to allow for location-independent measurements (Figure~\ref{fig:traditional-approach}), e.g., for international roaming measurements.
  Alternatively, SIM cards can be shipped around and switched manually.
  
  The first approach scales in $\mathcal{O}\left(|P|\times|S|\right)$, where $|P|$ is the number of probes, and $|S|$ is the number of SIMs.
  The second approach binds human resources at every location and limits the number of possible time-shared measurements.
  Additionally, SIM orchestration (i.e., physically provisioning SIMs to other locations) and provider mobile plans (e.g., one contract per location) are the highest cost- and maintenance drivers in a measurement system \cite{alay2015monroe}. %
  
Thus, a measurement system without flexible switching between SIM cards at target locations does not meet the requirements for large-scale cellular measurements. %

\item[\roundframe{P3}] \textit{Background noise.}~
  Some measurement frameworks lack exclusivity regarding the network interface.
  This is not a problem in many qualitative measurement settings (e.g., IPv6 capabilities); however, exclusivity is required for most quantitative measurements (e.g., how IPv6 is metered).
  Especially metering is often unavailable in real-time and requires careful coordination with other tests.
  
  Depending on the target, even a tiny amount of background traffic (e.g., DNS lookups) can influence the results and call for sterile access to the network interface and other cellular services.
\end{itemize}

\subsection{Goals and Requirements}\label{sec:requirements}

Based on these problems, we identified the following requirements for a new system:

\begin{itemize}[topsep=3pt plus 2pt,itemsep=0ex]\setlength{\itemsep}{3pt plus 2pt}\setlength{\parskip}{1pt plus 2pt minus 2pt}%

    \item[\roundframe{R1}] \textit{Scalability.}~
   A proper, scalable system must deal with a multi-dimensional combination space of different mobile providers, access technologies, carrier products, tariffs -- all of these in many different countries and with every roaming partner. Thus, manually managing these components is tedious and costly.
    
    The system should achieve complete coverage of all combinations without manual work, duplication of hardware setups, or multiple identical mobile plans.
    Security measurements need a flexible and easy-to-deploy base that enables security investigations of edge-cases, where unhabitual processes or actors are involved (e.g., roaming).
    Given the profound impact on the completeness and validity of results, we identify scalability as the essential requirement.

\item[\roundframe{R2}]\textit{Control of SIM communication.}~ 
    SIM cards are a significant component in the cellular network stack.
    These often underestimated microcontrollers can receive over-the-air (OTA) updates from the operator, load custom SIM applets, and proactively send commands to the modem they are connected with.
    Moreover, they contain the key material that is the baseline to generate session keys for radio and also VoLTE connections.
    Smartphone operating systems and their network stacks hide low-level SIM communication from the application layer.
    Without the ability to observe and possibly change communication, any measurement system lacks fundamental insights into the access network.
    
    To gain these insights and to enable security- and privacy-related auditing of the opaque communication stream between SIM card and modem, we identify control of SIM communication as one of our design goals.
    
\item[\roundframe{R3}]\textit{Utilization of the full feature spectrum.}~
  A cell phone connection is much more than just an Internet outlet. 
  The selection of specific operators, cells, and access technology shapes the properties of the data channel. 
  Additionally, voice, SMS, and USSD codes each form a complete ecosystem of their own. 
  Ignoring these features deprives the researcher of substantial tools to gain in-depth knowledge on the measured networks and may hide important aspects that may be valuable from a security and privacy perspective.
  
  Therefore, a new measurement system should provide full access to cellular-specific parameters and features.

\item[\roundframe{R4}]\textit{Isolated \& Controlled Environment.}~ 
    Many operator-related systems (e.g., billing, provisioning, QoE, and QoS) are opaque and thus difficult to analyze.
    A clean, noise-free (or -reduced) environment is paramount for accurate measurements and conclusions.
    Only application- and traffic-exclusivity with tight control over the network stack and possible background data will provide the required data quality.
    Additionally, only a geographically stationary system can provide the environment to gather stable and comparable longitudinal data.

    \item[\roundframe{R5}]\textit{Breadth.}~ 
	Networks and all their components are very complex. This also holds true for the interplay between different access technologies as well as networks, both of which can have unintended or unanticipated side effects.
    The handling of such corner cases is often not covered by standard procedures (e.g., billing), and, in some cases, not even the network operator might be aware of these effects.
    Root causes might be ad-hoc decisions during network design and setup or the fact that the operator lacks means to test traffic flow through all combinations of foreign networks.
    
    To enable security inspection of complex corner cases, a newly developed system should provide a broader overview than a single network operator has. %

\item[\roundframe{R6}]\textit{Low-cost, self-managing, open-design.}~ 
	The geographical and functional breadth can only be achieved through a simple deployment by an (easily recruitable) community of station caretakers.
    Neither they nor the overall organizers should need to invest a lot of resources to enhance, grow, or maintain the system.
    Furthermore, low-cost design and off-the-shelf hardware reduce the entry barrier for further research and may lower the barrier for testing known security risks in advance.
\end{itemize}

\section{System Design}\label{sec:system-design}

In this section, we first describe the key design elements and then provide a system overview.

\subsection{Geographical Decoupling of SIM Cards and Modems}\label{subsec:geographically-decoupling-of-sim-cards-and-modems}

A mobile device and its SIM are often viewed as an indivisible tandem, assuming that one cannot work without the other.
However, testing mobile connections through SIMs issued by different operators in different networks (e.g., roaming) includes (a) physically moving SIMs or (b) physically moving SIMs and their mobile devices, or (c) replicating each SIM and the measurement setup in every territory (cf. Figure~\ref{fig:traditional-approach}) --- thus, scaling poorly based on manual or financial effort.

There are numerous ways to geographically decouple the identity (i.e., SIM) from its actual usage in a particular network.
At higher layers, the mobile device's  radio stack communicates via a low-bandwidth protocol with the SIM.
Moving down through the stack, it produces radio messages in the lower layers until it modulates high-bandwidth, low-latency digital radio samples.
At the lower layers, it transforms these samples into an analog radio frequency (RF) signal and amplifies it before radiating it via an antenna (and vice versa for receiving).
Thus, a geographical displacement of the lower layers entails a much higher engineering overhead than at higher layers. 
While software-defined radio (SDR) would also provide more freedom on the radio layer (i.e., custom or non-standard messages) and open the system to air interface testing, their use is generally not permitted on real-world networks. Even though strong GSM, LTE, and 5G software implementations for SDRs exist, they lack the radio-regulatory permissions to operate.
Therefore, using a globally-licensed off-the-shelf modem and decoupling on the low-bandwidth communication between the SIM and the modem provides the best trade-off for rapid global deployment. 

Based on these considerations, we opted for tunneling the SIM card's ISO7816 T=0 protocol~\cite{iso7816-3} through the Internet by electrically connecting the modem's SIM socket with GPIO pins of our measurement platform, and recreated the protocol via software. Modem and SIM exchange Application Protocol Data Units (APDUs) via the T=0 connection. %

Connecting any SIM with any radio module (i.e., modem) -- regardless of geography -- facilitates measurements across a large number of SIMs and radio networks.
SIMs can be either pooled centrally or hosted de-centrally, using an infrastructure to add and remove SIMs dynamically.

This method fulfills the scalability \roundframe{R1}, low-level control \roundframe{R2}, and breadth \roundframe{R5} requirements.
Using GPIO pins directly and implementing T=0 in software reduced costs and complexity \roundframe{R6}. 
Furthermore, in our design, we do not have to support all protocol speeds and electrical variants on both ends of the tunnel as they are electrically independent. %

\paragraph{eSIM}
eSIM is a complementary technology that overlaps slightly with to our proposed technique but can not replace it for two reasons. First, eSIMs are not available from all operators, and where available, they are not always obtainable for all (subscription) plans. Second, eSIM deployment is not always device-independent and may require different distribution methods. For example, some eSIMs are only available via specific apps -- some of which are not (virtual) MNOs but merely resellers. Others require network support. Some operators only make them available for specific device types (e.g., Apple smartwatches). Furthermore, the eSIM ecosystem is heavily fenced off with certificates.  
Thus, our solution is the most universal, as we can tunnel physical SIMs as well as eSIMs from supported devices via Bluetooth (remote) SIM access profile (rSAP), as presented in Section \ref{subsubsec:impl-bluetoth-sim-access}.

\subsection{Process and Network Isolation}
As we intend to reduce network noise \roundframe{R4}, we can not use ordinary smartphones - as the operating system would induce too much background traffic noise. However, even in our dedicated system, we needed to split the network stacks.  
The SIM tunnel and control connections need to be separated from the measurement system.
Docker and other container systems use Linux network namespaces~\cite{docker-linux-namespaces} to create shielded networking enclaves. Such a network namespace has a separate device configuration and separate routing tables.
In these setups, only selected processes are admitted to a namespace that uses the cellular connection as their default gateway, thus, shielding the measurement script from any other system activity. %

\subsection{Metering Measurements}\label{subsec:metering-measurements}
With most network operators, billing is not a real-time service.
Charge Data Records (CDRs) are collected decentrally and accumulated asynchronously in the customer records.
Delays span from minutes to hours, depending on the operator.
While some operators account for every single byte, others collate (and usually round up) the charges.
This may also vary in roaming situations since additional parties are involved in the accounting process.

For customers (and researchers), there is no standardized way to query data plan usage.
Most operators offer a custom app or website. In addition, some providers still operate USSD and SMS methods as used in legacy SIM Toolkit applications. This is captured by the \roundframe{R3} feature utilization requirement.

In order to speed up the measurement of different traffic classes, we can utilize a binary power encoding scheme by varying the data amount, i.e., $\mathit{amount}=2^{\mathit{class\,nr}}$.
For example, one MB of class 1 is produced, two MB of class 2, four MB of class 3, and so on.
Hours later, when the CDRs are available, it is possible to separate these traffic classes.

As metering measurements are one of the possible use cases of a new cellular measurement system, built-in credit-checking support is advisable.

\subsection{Overview}\label{subsec:component-overview}

Our measurement framework is designed as follows.

\begin{figure}[!t]%
    \centering
    \includegraphics[width=1\linewidth]{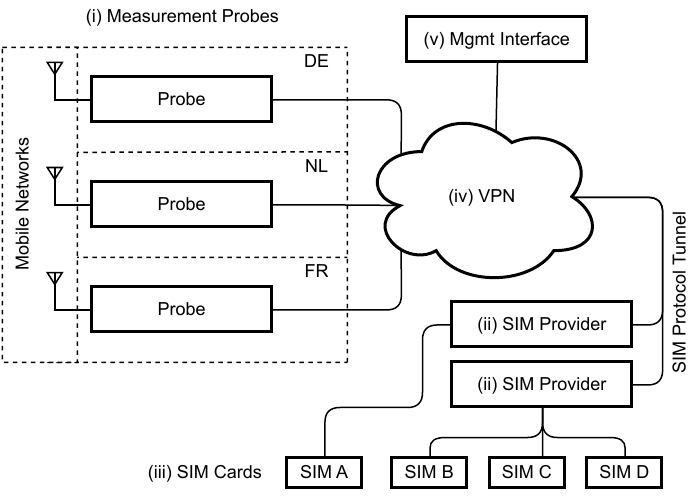}
    \caption{High-level overview of \toolname components: SIMs connected to our SIM Providers can be virtually connected to any measurement probe at the target location.}
    \label{fig:architecture}
\end{figure}

Our fundamental architecture consists of five components, as illustrated in Figure~\ref{fig:architecture}:
\begin{enumerate}[topsep=0pt plus 1pt,itemsep=0ex]\setlength{\topsep}{0ex}\setlength{\itemsep}{0ex}\setlength{\parsep}{0ex}\setlength{\parskip}{0pt plus 3pt}
    \item[(i)] Probes that execute experiments and measurements on the cellular network via a modem.
               Furthermore, the probe electrically emulates the SIM card for the modem.  
    \item[(ii)] SIM providers that connect to multiple genuine operator SIM cards (i.e., concentrate them or allow users to \textit{temporarily donate} a SIM with special properties), allowing to connect them to any modem via a virtual circuit.
    \item[(iii)] The actual SIM cards, which contain the secret key from the operators.
    \item[(iv)] A separate VPN-protected Internet connection for management purposes, allowing deployments behind NATs and firewalls.
    \item[(v)] A management interface to monitor and administrate the components and orchestrate tests.
     It remotely updates and introduces new measurement scripts. Additionally, it collects the status (dashboard) and allows interaction with probes (e.g., schedule tasks, connect to selected SIMs, reboot probes).
\end{enumerate}

\subsection{Ethical Considerations}\label{subsec:ethical-considerations}

We consider ethical aspects for three roles: The designer of the method, the operator of the testbed and measurement network, and the researcher that executes tests and measurements. 
Not all of the following apply to all of the roles.

\paragraph{Technical and radio regulatory.} 
Radio regulatory compliance and certification is paramount if stations are to be operated by third parties or lay people. 
Therefore we can not use SDRs, and we only employ globally-certified modems and refrain from making any (hardware) modifications to them.  
We only capture and transmit signals (e.g., APDUs) via the designated external interfaces.

\paragraph{Live network influence.}
We conducted our measurements in real-world, live mobile networks;
our metrics reflect normal user behavior (data access, SMS functions, and telephone calls). Ideally, our activities should not affect the network.
However, due to SIM tunneling, our SIM cards change the ``country'' in an irregular and fast fashion.
This might spark confusion among the operator systems or trigger fraud control alerts.
In order to exercise caution, we manually impose a waiting time between country switches.

\paragraph{Guest network access.}
Our probes are connected to our host's LANs.
We hardened the operating system configuration and executed measurements exclusively in their own namespaces.
To further facilitate secure deployment, we ensured that the Internet-facing control traffic passes to the control server only via our VPN.

\paragraph{SIM registration.}
Regarding the use of SIM cards, we must consider %
that SIM registration laws in certain countries may require personalization or registration of SIM cards~\cite{gsma2013mandatory, gsma2019access}.
Where required, we registered the SIM cards to ourselves and exclusively used them for measurements.

\paragraph{Provider's terms of service.}
Probing, testing, and speed measurements could violate an operator's terms and conditions.
We took precautions not to harm the networks and did not use our technology to exploit networks economically. Furthermore, we share the implementation and results, as this is the only way -- in our opinion -- to ascertain robust public infrastructure knowledge.
While part of the project shares similarities with SIM banks, we do not conduct any of the other activities of the SIM bank fraud scheme, such as terminating calls on local SIM cards, faking International Mobile Equipment Identifiers (IMEI), or enriching ourselves.

\paragraph{Vulnerability exposure.}
Measurements aim to improve understanding of cellular networks and their interplay. 
Whenever we identify serious problems leading to harmful behavior, we contact the network operators directly and report our findings responsibly prior to publication.

\paragraph{Economic losses.}
After careful evaluation we are convinced that there is no other way of testing real-world billing systems. 
Thus, we always make sure not to enrich ourselves by not oversizing the generated traffic and by letting an equal or greater amount of our monthly traffic allowance expire at the end of the month (as if we were billed for the traffic).

\section{Implementation}\label{sec:implementation}

This section is a brief summary of the implementation details and choices as well es some performance measurements described in the Appendix.%

\subsection{Probe}
\paragraph{Probe Hardware} 
The hardware went through two major iterations for cost, stability, and supported features -- as it was revised after the deployment of the first batch. Since a large-scale deployment requires a grassroots-like approach and costs multiply by the number of probes, we eventually settled for a low-cost \textit{Raspberry Pi 4} and a \textit{Quectel EG25-G} modem via a mini-PCIe adapter. The SIM is controlled via an UART on GPIO pins, and the Internet connection is realized via the Ethernet port. 

\paragraph{Probe Software Architecture}
The architecture consists of a Python-written dynamically deployed test-suite that runs in a Linux (network namespace) container for isolation, similar to Docker. Only the containers traffic is routed through the modem and recorded to a PCAP file. 

\subsection{SIM Interface and tunneling}
\paragraph{SIM Interface}
Multiple generations of the synchronous SIM card interfaces in various voltages (1.8 V, 3 V, and 5V) with multiple speed (clock speeds and clock divider) settings exist. Our approach is significantly simplified in contrast to other SIM-interfacing tools (e.g., SIM trace\cite{simtrace}) in that both ends of the tunnel are electrically separated. We can thus implement and negotiate speed and voltage independently through the \textit{Answer on Reset} (ATR) handshake and the \textit{Protocol and Parameter Selection} (PPS) command. Since the on-board UART only supports asynchronous transfers, we approximate the speed characteristics. Generous divider settings, tolerances, and the ability to start communication at any clock cycle work in favor of an extreme minimalistic hardware interface. For the open-collector data bus, the modem already provides a pull-up, and the UART driver pin is shielded via a Shottky diode (Figure \ref{fig:sim-interface}). 

Additionally, the (simulated) SIM has the ability to signal \textit{Waiting Time eXtensions} (WTX) towards the modem to mask the effects of network latency.

\paragraph{SIM Provider}
The dynamic SIM tunnel eventually has to terminate at a real (e)SIM. 
For a medium-sized operation, cheap USB card readers (either via a serial or PC/SC interface) are the most economic option. Dedicated hardware (SIM Banks), sometimes used for OTT-Fraud\cite{sahin2016over}, serves more SIM cards but lacks standardized interfaces.
While USB theoretically supports up to 127 devices, in our experience we seldom observed a reliable operation beyond 32 devices even on quality hardware.

\paragraph{eSIM and Bluetooth SIM Access Profile} 
We have also implemented a Bluetooth-based SIM provider via the \textit{remote SIM Access Profile} (rSAP), such as used by many cars. 
This offers a route to attach \textit{embedded SIM} (eSIM) into the \toolname system: 
Off-the-shelf non-rooted Android phones (in our example a Google Pixel 3a) are happy to share an eSIM via rSAP when it is configured as primary SIM.

\subsection{Deployment}\label{subsec:deployment}\label{sec:deployment}

We currently have probes in eight European countries (Austria, Belgium, Croatia, Finland, Germany, Romania, Slovenia, and Slowakia), and two North American countries (Canada and the United States).
These ten countries account for more than 510 million people.

We argue that for core network, subscriber billing, and roaming tests, one probe with good radio connectivity per country is sufficient. These are operator-wide properties (Figure \ref{fig:mnostruct}), not dependent on individual base stations.
Some low-level or fine-grained RAN testing might need multiple stations, such as testing handovers, radio propagation under different geographical conditions, domestic roaming (as used in China until 2018), or special radio installations (e.g., cruise ships).

We included SIM cards from bare-metal (MNO) and virtual network operators (MVNO) in our Proof-of-concept setup. eSIMs are supported by via Bluetooth's (remote) SIM Access Profile (SAP). %

\subsection{Benchmarks}
A greater variety of benchmarks can be found in the Appendix \ref{subsec:benchmarking}. The tunnel latency stands out in its importance for deployment.
In a local setting, APDU round-trips are mostly negligible.  
However, it was never designed for a remote-tunnel setting where latency and round-trip numbers multiply.

\subsubsection{SIM tunneling latency analysis}\label{subsubsec:sim-tunneling-latency-analysis}\label{sec:impl_benchmark_tunnellatency}

Tunneling inevitably increases latency at the SIM interface and subsequently in the network (e.g., for authentication and session key generation round-trips).
As described above, we used WTX to avoid timeouts from the modem.
For this section, we tested the tunnel's robustness on the radio interface to the network operators by artificially inserting long %
delays.

We tested three different operators with the Quectel EG25-G and added an artificial delay of 1,000 ms before each APDU response.
This matches high-latency network conditions.

In all cases, the modem successfully attached to the cellular network.
We verified the functionality by sending and receiving SMS and initiating a voice call.

Once the modem is attached, APDU traffic is minimal. %
We could not detect any network performance loss (e.g., data, SMS, call) due to the SIM tunnel. In general, handovers between cell towers do not cause re-authentications. However, simultaneous radio access technology changes (e.g,. LTE to UMTS) can trigger re-authentications and cause a slight delay.

Operators are aware of vastly diverging authentication round-trip times~\cite{Dabrowski2016raid} between devices.
However, in theory, our extreme round-trip times could stand out. %

\section{Showcases}\label{sec:showcases}

This section presents five security and privacy focused use cases demonstrating diverse capabilities of \toolname.
The first group of use cases demonstrates a few of the vast possibilities for roaming-related measurements and studies.
The second group showcases how to test a hypothesis, measure a network property, or collect provider configurations on a large scale domestically and internationally.

\subsection{Roaming and Price Differentiation}
\label{subsec:roaming-and-price-differentiation}
\label{sec:showcase-roaming-and-price-differentiation}

Previous work on differentiation~\cite{dischinger2010glasnost,li2019large,bashko2013bonafide,zhang2009detecting, kakhki2016bingeon} analyzed traffic as well as the implications of international data roaming~\cite{mandalari2018experience}.
However, the question arises if zero-rated traffic is handled differently in roaming.
Understanding roaming billing is essential as it is either much more expensive than domestic traffic or subject to stricter limits (as in the EU's free-roaming agreement \cite{eu-reg-2007-717}).
We hypothesized that the roamed traffic classification is identical due to home routing; however, unexpected differences in metering exist.
In this use case, we focus on the measurement of fine-grained metering differentiation of roamed DNS traffic, 
which would inform an attacker how and when to use DNS tunnels to hide traffic from billing.

\begin{table}[!b]%
\definecolor{highlight}{RGB}{255,229,229}
\centering
\caption{DNS billing for 14 providers from five countries. P-RO-1 and P-SI-1 differ in roaming billing behavior.}\label{tab:dns-results}
\scalebox{0.85}{%
\begin{tabular}{|p{2cm}p{2cm}X{1.5cm}X{1.5cm}|}
\hline
& & \multicolumn{2}{c|}{Billing} \\
Provider & DNS IP & Domestic & Roaming\textsuperscript{\textdagger}  \\
\hline
P-AT-1       &  private  &  \LEFTcircle    &  \LEFTcircle   \\
P-AT-2       &  public  &  \LEFTcircle    &  \LEFTcircle   \\
P-AT-3       &  public  &  \LEFTcircle    &  \LEFTcircle   \\
P-HR-1       &  public  &  \LEFTcircle    &  \LEFTcircle   \\
P-HR-2       &  public  &  \LEFTcircle    &  \LEFTcircle   \\
P-HR-3       &  public  &  \LEFTcircle    &  \LEFTcircle   \\
P-RO-1       &  public  & \cellcolor{highlight} \LEFTcircle    &  \cellcolor{highlight}\CIRCLE         \\
P-RO-2       &  public  &  \CIRCLE          &  \CIRCLE         \\
P-RO-3       &  public  &  \LEFTcircle    &  \LEFTcircle   \\
P-SK-1       &  public  &  \CIRCLE          &  \CIRCLE         \\
P-SK-2       &  public  &  \LEFTcircle    &  \LEFTcircle   \\
P-SK-3       &  private  &  \CIRCLE          &  \CIRCLE         \\
P-SI-1       &  public  & \cellcolor{highlight} \LEFTcircle    &  \cellcolor{highlight} \CIRCLE         \\
P-SI-2       &  public  &  \LEFTcircle    &  \LEFTcircle   \\
\hline
\end{tabular}%
} %
\begin{minipage}[l]{0.85\linewidth}\scriptsize
\vspace{1mm}
\CIRCLE~Internal \& external DNS traffic billed \\
\LEFTcircle~Only external DNS traffic billed \hfill \smash{\colorbox{highlight}{discrepancy}}\\
\Circle~DNS traffic not billed \ \\
\textsuperscript{\textdagger}~Measured with automatic network selection in two different EU countries
\end{minipage}%
\end{table}

\paragraph{Methodology.}
We tested the provider's default DNS configured via DHCP (internal), as well as an external self-hosted DNS resolver. %
First, the current data quota is checked automatically (see Section~\ref{subsec:metering-measurements}).
Second, we repeatedly sent DNS queries (for top domains according to Tranco~\cite{LePochat2019} list V78N) to the target DNS server until the necessary traffic size (e.g., one MB) was reached.
Third, we waited for the billing systems to update and recheck the available quota.
This process was repeated for each SIM card, each tested country, and each tested DNS server.
Finally, we analyzed the metering differences between domestic and roaming data usage. %

\paragraph{Results.} 
Table~\ref{tab:dns-results} presents the results as of March 2022.
External DNS traffic was always billed by all providers in both scenarios (i.e., domestic and roaming).
Nine providers did not bill internal DNS traffic regardless of roaming, whereas three others did.  
Interestingly, two providers (P-RO-1, P-SI-1) behaved differently under roaming conditions: Internal DNS traffic is not billed in the home network but is subtracted from the data quota when abroad.
Twelve providers host their customer-facing DNS servers (that are served via DHCP) in public IP ranges; the other two have private IPs (P-AT-1, P-SK-3).
Furthermore, we noticed that the DNS queries took noticeably longer during roaming due to increased round-trip times within home-routed connections (cf. Section \ref{subsec:network-and-firewall-configuration}).

These results demonstrate that roaming differentiation in billing exists and that \toolname can be utilized for large-scale and fine-grained analysis.

\subsection{Zero-Rating and Free-Riding}
\label{subsec:zero-rating-and-free-riding}
\label{sec:showcase-zero-rating-and-free-riding}

Zero-Rating specific applications (e.g., Facebook, Instagram, Snapchat) within dedicated data packages or tariffs is a common practice for many operators to reach specific market demographics.
To correctly exempt traffic of those applications from a customer's data quota, all incurring data packets need to be classified by the billing system.
For maintenance reasons, many providers do not use IP-based whitelisting because those can change on short notice in cloud-hosted environments.
In early experiments, we discovered that many operators revert to client-supplied HTTP's hostname and TLS's SNI headers instead.  %
This opens the door for \textit{phreaking} or \textit{free riding}, where a user can disguise traffic so it is misclassified by the billing system, allowing them to dodge fees.

\paragraph{Methodology.}
We chose Snapchat as one of the most popular zero-rated services for our experiment. We tested all operators for which we could obtain a matching subscription plan or data package (see Table~\ref{tab:dns-results}).
First, we validated that the zero-rating package is active by sending web requests to a web endpoint that is used within our target application (i.e., app.snapchat.com).

Second, we evaluated whether the host header is used for classification and can be abused for free-riding.
Instead of connecting to the legitimate web endpoint, we make the same request to an AWS server under our control that accepts web connections for arbitrary host and SNI names.
We run each test for HTTP and HTTPS in a domestic and roamed usage scenario.
After analyzing the billed data traffic, we know whether the legitimate application is zero-rated and which operators are vulnerable to free-riding by a spoofed hostname.

\paragraph{Results.}
The results for May 2022 are presented in Table~\ref{tab:free-riding-results}.
While we verified Snapchat zero-rating during domestic usage for all tested providers, one provider (P-HR-1) is still billing their roaming customers.
Three providers turned out to be vulnerable to spoofed host headers.
In practice, an attacker could abuse this type of classification by sending arbitrary data via a proxy server masquerading as the same host.
Although the legitimate application communicates exclusively via HTTPS, two providers also zero-rate legacy HTTP traffic with a matching hostname header.
The one provider in our sample that was not vulnerable to our free-riding attack via spoofed host and SNI names presumably uses a different metric (e.g., IP-ranges) for traffic classification.

\subsection{Network- and Firewall-Configuration}
\label{subsec:network-and-firewall-configuration}
\label{sec:showcase-network-and-firewall-configuration}

\toolname is a great tool for quickly determining IP level configurations for different operators or usage scenarios.
This includes observation of used IP address ranges, as well as experiments that test for deployed firewall rules.
Previous work has shown that security configurations can be different for IPv4 and IPv6 on dual-stack enabled servers and routers~\cite{czyz2016don}.
\toolname supports IPv4, IPv6, and dual-stack connections, which makes it an excellent tool to investigate, whether this assumption holds for the cellular field.

For roaming, there are two common scenarios: local breakout (LBO) and home routing (HR) (see Section \ref{sec:background_roaming}).
When LBO is used, the customer gets an IP address from the roaming partner. 
With HR, all data is forwarded to the home network. %
While HR usually provides better security, it often adds additional latency because all packets are routed via the home network.

\begin{table}[!t] %
\centering
\caption{With many providers, zero-rating can be abused for free-riding.}\label{tab:free-riding-results}
\resizebox{0.9\linewidth}{!}{%
\begin{tabular}{@{}l|cc|cc@{}}
& \multicolumn{2}{c|}{Zero-Rating} & \multicolumn{2}{c}{Free-Riding} \\
Provider & Domestic & Roaming & HTTP(Host) & HTTPS(SNI) \\
\hline
P-AT-1 & \checkmark & \checkmark     & \checkmark  & \checkmark  \\
P-AT-2 & \checkmark & \checkmark     & $\times$    & $\times$    \\
P-HR-1 & \checkmark & $\times$       & \checkmark  & \checkmark  \\
P-HR-2 & \checkmark & \checkmark     & $\times$    & \checkmark  \\
\end{tabular}%
}%
\end{table}

\paragraph{Methodology.}
Since \toolname provides verbose insights into the parameters used for cellular network connections, we leveraged the data from our previous use cases to analyze these configurations.
For all operators that provide us with a public IP address, we ran additional tests (e.g., ping and a low-volume nmap scan) against our own device to check whether incoming connections are filtered. 
(Note: No other customers were scanned.)

\paragraph{Results.}
All 14 tested providers use carrier-grade NAT for IPv4, and only two (P-AT-1, P-AT-2) offered IPv6 addresses (single- or dual-stack).
The provided IP range for the carrier-grade NAT was 100.64.0.0/10 for three providers (P-SK-3, P-SI-1, P-SI-2), 100.64.0.0/10 + 10.0.0.0/8 for two providers (P-AT-2, P-HR-2), and just 10.0.0.0/8 for all other providers\footnote{According to RFC 6598~\cite{rfc6598}, only 100.64.0.0/10 should be used for Carrier-Grade NAT to prevent IP address conflicts with private ranges.}.
During our intra-EU roaming measurements, all 14 tested providers used home-routed data connections.
Therefore the deployed IP configuration was identical in both domestic and roaming usage scenarios.

All used IPv4 addresses lie within non-public ranges; thus, the carrier-grade NAT acted like an implicit firewall against incoming connections.

For IPv6, however, both providers (P-AT-1, P-AT-2) use public IP ranges.
P-AT-1 is blocking incoming connections, allowing the feature to be switched off in the customer profile. 
However, P-AT-2 is fully accessible from the outside and provides no such option. 
This might reveal application service ports to the public Internet and surprise many users and developers who rely on the implicit firewall~\cite{czyz2016don, Ullrich2014ipv6sok}.
Several applications knowingly or accidentally expose TCP ports to the cell-network facing side, e.g., Netflix with port 9080, and the Android Debug Bridge (adb) on port 5555 (when \textit{tcpip} mode is enabled).
Apart from that, low-level attacks with traffic-usage inflation are possible due to the public IP address.\looseness=-1

\subsection{Low-level APDU analytics}
\label{subsec:low-level-apdu-analytics}
\label{sec:showcase-low-level-apdu-analytics}

\begin{table}[!t]%
\centering%
\caption{Example of a proactive binary SMS by P-AT-2. The opaque message contains encoded IMSI, ICCID and IMEI of the used devices.}\label{table:binary-sms}
\renewcommand{\arraystretch}{1.2}%
\setlength{\tabcolsep}{2pt}%
\def\g{\cellcolor{black!15}}%
\newcommand{\sep}[3]{\multicolumn{1}{#1>{\tt}c#2}{#3}}%
\newcommand{\desc}[3]{\multicolumn{#1}{#2}{\smash{\raisebox{-1pt}{\tiny #3}}}}%
\newcommand{\numbering}[2]{\multicolumn{15}{l}{\raisebox{5pt}{\tiny #1}}&\multicolumn{1}{r}{\raisebox{4pt}{\tiny #2}} \\[-20pt]}%
\scalebox{0.85}{%
\begin{tabular}{|>{\small\tt}c>{\small\tt}c>{\small\tt}c>{\small\tt}c>{\small\tt}c>{\small\tt}c>{\small\tt}c>{\small\tt}c>{\small\tt}c>{\small\tt}c>{\small\tt}c>{\small\tt}c>{\small\tt}c>{\small\tt}c>{\small\tt}c>{\small\tt}c|}
\hline
05&33&ff&81&81&81&81&81&81&82&ff&ff&ff&ff&ff&ff\\
\hline \numbering{0}{15} \desc{16}{c}{} \\
\hline
ff&ff&ff&ff&ff&ff&ff&ff&ff&ff&ff&ff&ff&ff&ff&ff\\
\hline \numbering{16}{31} \desc{16}{c}{} \\
\hline
01&01&81&81&81&01&81&01&01&02&ff&ff&ff&ff&ff&ff\\
\hline \numbering{32}{47} \desc{16}{c}{} \\
\hline
ff&ff&ff&05&ff&08&01&42&07&02&03&12&24&0a&f7&de\\
\hline \numbering{48}{63} \desc{16}{c}{} \\
\hline
1f&9c&a7&9e&1f&e2&c3&11 & \sep{|}{}{\g 62} &\g 09 &\g 83 &\g 76 &\g 96 &\g 08 &\g 54 &\g 93\\
\hline \numbering{64}{79} \desc{8}{c}{} & \desc{8}{c}{IMEI SV$\rightarrow$} \\
\hline
\g 96 &\g 06 & \sep{}{||}{\g f8} & \g 01 & \g 0a & \g 98 & \g 34 & \g 30 & \g 00 & \g 00 & \g 12 & \g 33 & \g 03 & \g 53 & \sep{}{|}{\g 90} & 02 \\
\hline \numbering{80}{95} \desc{3}{r}{IMEI SV} & \desc{12}{c}{ICCID} \\
\hline
0a&07&e4&41&01&00&00&00&d0&03&a1&04&02&01&ff& \sep||{\g 0a}\\
\hline \numbering{96}{111} \desc{15}{r}{} & \desc{1}{c}{} \\
\cline{1-10} 
\g 09 &\g 08 &\g 29 &\g 23 &\g 30 &\g 03 &\g 12 &\g 41 &\g 52 & \sep{}|{\g 07}  \\
\cline{1-10} \numbering{112}{} \desc{12}{c}{$\rightarrow$IMSI} \\
\end{tabular}%
} %
\end{table}

In 2019, the Simjacker~\cite{simjacker} report gained broad media attention as it found that many SIM cards were not protected against malicious over-the-air updates, given that many operators did not implement authentication. %
In 2021, McDaid~\cite{mcdaid2021stk} showed that this is still an issue in many countries.

Hidden SIM communication between the operator and the SIM can happen in two ways: It can be initiated by (i) the operator through a binary SMS message (e.g., over-the-air update) or (ii) proactively by an application on the SIM at any time.
The latter can send binary SMS as well as instruct the phone to open TCP connections~\cite{proactive-sim1, proactive-sim2}.
Although playing a crucial role in the security of cellular networks, the operator's SIM card is a black box, demonstrating why more research is needed in this area.

\toolname is an excellent research platform for this purpose because it provides complete insight into the hidden communication channel,
especially if the SIM breaks the usual pattern of just answering the operator and starts proactively sending out requests.

\paragraph{Methodology.}
By default, \toolname stores all APDU traffic in a database. Therefore we were able to investigate recordings from our previous use cases for this analysis.

\paragraph{Results.}
We observed two interesting cases of hidden proactive SIM communication.
Two of the 14 SIM cards (P-AT-2, P-RO-1) sent SMS to the home operator without indicating this on the screen. %
Even though both were in an unknown binary format, we could clearly identify IMSI, ICCID, and IMEI in those messages.
An example from P-AT-2 is shown in Table~\ref{table:binary-sms}.
The remaining unidentified data fields was static over all observed messages.

\paragraph{Charges.} Even though those SMS were created outside the user's control by a device provided by the operator (i.e., the SIM), we were charged for them outside the EU's free-roaming area.

\subsection{Country-grained Location Leakage via Call Progress Tones}
\label{sec:showcase-location-leakage}

In modern telephony networks, signaling and audio are two separate channels (i.e., out-of-band). 
However, it became common (``early media") to establish the audio channels already during dialing (as opposed to establishing them when the call is accepted) and let the closest switching center to the target create the ringback, busy tone, or voice announcements. This concept allows for additional audio messages or ringback music to be played to the caller.
While recommendations for standard call progress signals exist by ETSI and other regulatory bodies, it is ultimately the operator to decide and the switching center on the callee side to create those signals with all the subtle differences that this might entail. Historically, the US uses dual-frequency ringback tones, while European networks use a single-frequency tone. However, even within Europe, subtle differences in on-off times, the primary frequency and side lobes, as well as the amplitude exist.

On a technical level, circuit-switched voice roaming is realized as a local breakout (see Section \ref{sec:background_roaming}) with a call forward to a temporary phone number in the visiting network, i.e., the visiting network creates the ringback or busy tones. 
In contrast, VoLTE can be broken out locally or home routed.

Previous work has already shown that sophisticated audio fingerprinting can be used to detect fraud (i.e., hijacked calls)~\cite{reaves2015boxed, peeters2018sonar} and to determine call provenance~\cite{balasubramaniyan2010pindr0p}.
We suspect it is possible in many cases to locate a cell phone subscriber's current network or country, based on simple metrics in the ringback tone of a single test call. Since MNOs are bound to a specific territory by the frequency regulatory bodies, one can infer the coarse location of a phone if the network operator is to be identified, especially in an area of the world with many small countries and permeable borders. This has various privacy implications. 

For example, a thief can test if a possible victim is abroad. A competitor could test if a high-level official of another company or country is in a specific country (e.g., for a specific summit). Alternatively, a nosy supervisor could test if an employee misuses home office or sick leave. 
This technique also allows the identification of the actual operator of a domestically ported cellphone number (e.g., in a SIM-swap attack).

\paragraph{Methodology.}
Since the audio signal is a digitally compressed audio stream, we assume that (under normal conditions) it is passed unaltered and the subtle differences originate in the target network.

In the first step, we record three voice call setups from one base point to a roaming target. Then the target SIM was set into a different roaming network, and the process repeated. If inconclusive, we added additional test calls. We used SIMs from three different operators from one country to virtually \textit{travel} to Austria, Canada, Croatia, Finland, Germany,  Romania, Slovakia, Slovenia, and the United States via 19 distinct roaming partners.

For tone analysis, we cut off recordings after 20 seconds to ignore the following voicemail greeting. %
Then we identified metrics and features to distinguish the ringback tones. 
Where available, we repeated the setup for VoLTE connections. 

We analyzed the data for call setup times, ringback duty cycle (on/off times), ringback signal amplitude(s), and ringback frequency composition, but also manually heard them to find any other anomalies.

\begin{figure}[!b]%
    \begin{subfigure}[b]{\linewidth}
        \centering
        \includegraphics[width=.8\linewidth,clip=true,trim=18px 210px 81px 50px]{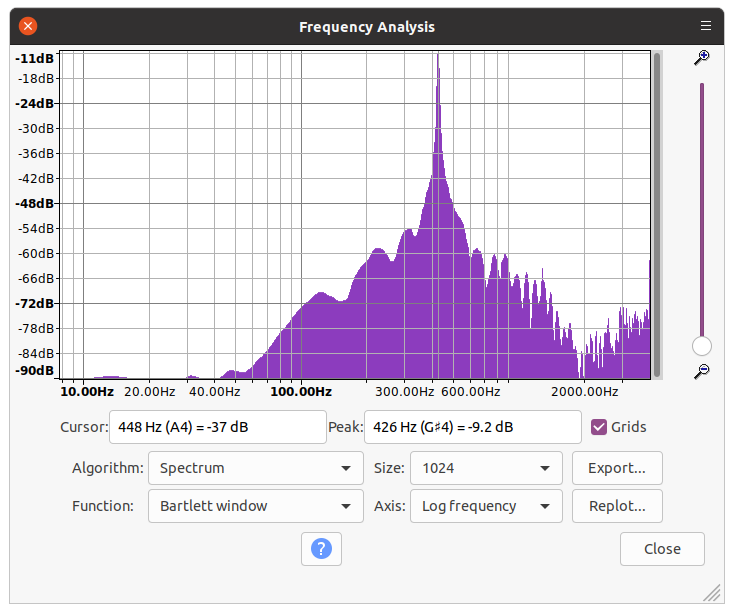}
        \caption{DE-o2's ringback tone is a loud 426~kHz sine wave with -9.2~dB.}
        \label{fig:spectrum_de2}
    \end{subfigure}

    \begin{subfigure}[b]{\linewidth}
        \centering
        \includegraphics[width=.8\linewidth,clip=true,trim=18px 210px 81px 50px]{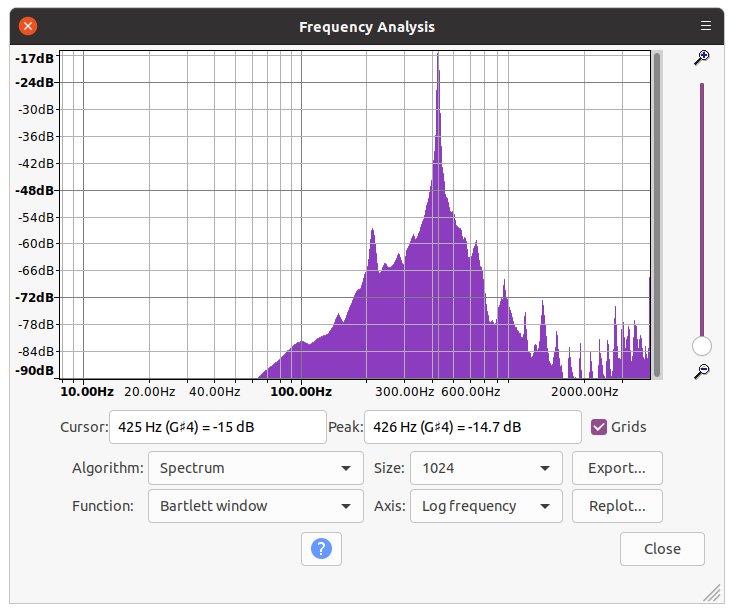}
        \caption{DE-Telekom's ringback tone at 426~kHz is quieter with a -14.7~dB signal with a side peak at 212~kHz.}
        \label{fig:spectrum_de1}
    \end{subfigure}

    \begin{subfigure}[b]{\linewidth}
        \centering
        \includegraphics[width=.8\linewidth,clip=true,trim=18px 210px 81px 50px]{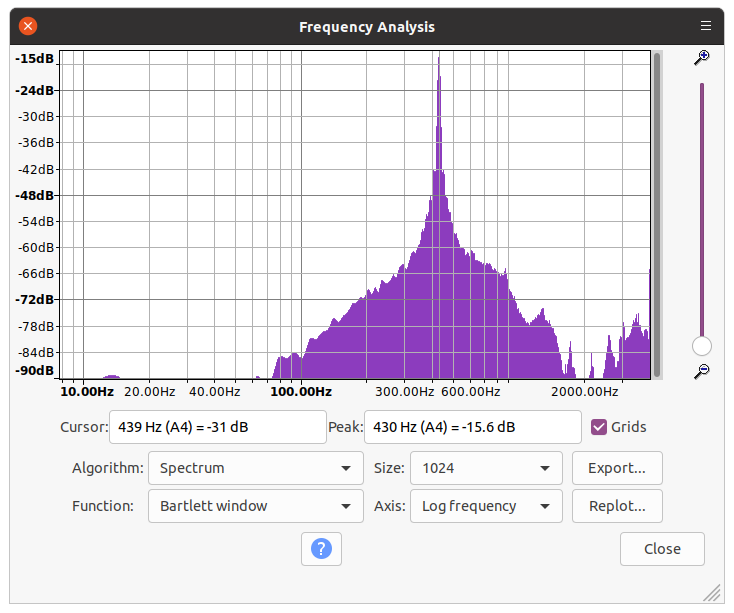}
        \caption{A clean RO-Vodafone ringback tone at 430~kHz with -15.6~dB.}
        \label{fig:spectrum_ro3}
    \end{subfigure}%
    \vspace{-2mm}
    \caption{Comparison of ringback spectra. }%
    \vspace{-2mm}%
\end{figure}

\begin{figure}[!t]%
    \centering
    \includegraphics[scale=0.31,clip=true,trim=30mm 15mm 182mm 30mm]{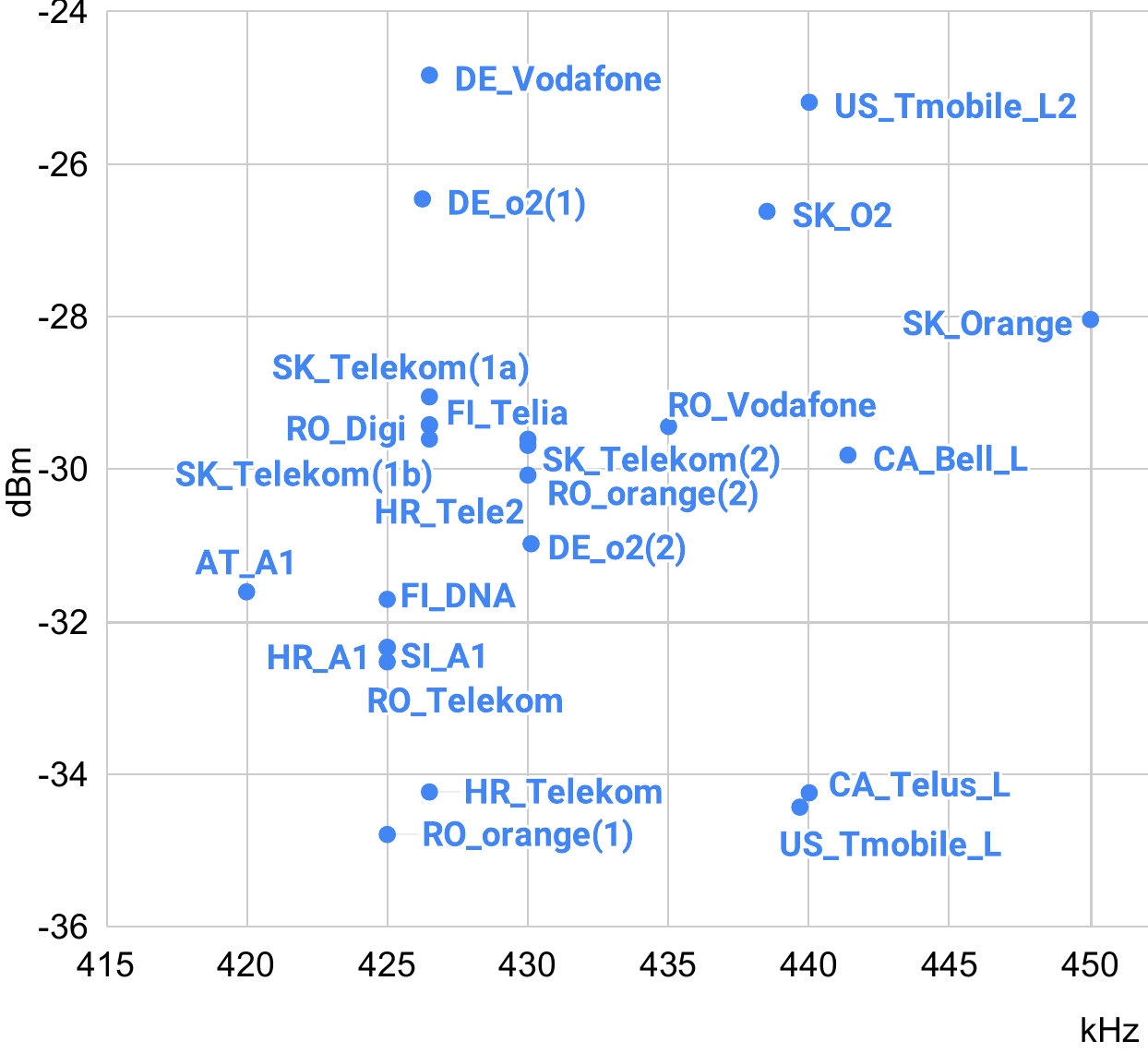}~\color{lightgray}\vline~%
    \includegraphics[scale=0.31,clip=true,trim=350mm 15mm 5mm 30mm]{images/callprogresstones-scatterplot.pdf}
    \caption{Fingerprinting ringback tones (without VoLTE).}
    \label{fig:ringback-fingerprint}
\end{figure}

\paragraph{Results.}

\textit{Early media} call progress tones heavily depend on the combination of source and (foreign) target network and the used access technology. Except for a few combinations, most circuit-switched voice calls (CSFB on LTE, 3G, and GSM) used target-network-generated tones with stable results. However, some networks seem to have two distinct settings, e.g., from a misconfigured load-balancing setup.
 VoLTE more often used origin-generated ringback tones. 
On average, VoLTE also had the shortest call setup times, and LTE's CSFB was the slowest (as it has to switch to 3G/2G first).

As for the ringback tone itself, we found several distinguishing features: (i) the base frequency, (ii) over-tone composition, (iii) on-off duty times, and (iv) the signal amplitude. 
The repeated recordings enforce our confidence in the stability of these features.

Call setup times turn out to be less stable and are likely influenced heavily by outside factors.
Furthermore, we noticed transient audio channel artifacts with some operators when the call is transferred to voice mail. 

\begin{enumerate}[topsep=2pt plus 2pt,itemsep=0ex]\setlength{\itemsep}{0ex}\setlength{\parskip}{0pt}
\item[(i)] \textbf{Base frequency.} In our European sample, the most prominent base frequencies are 420, 425, 426(.5), and 430 kHz. In contrast, all tested North American networks use a dual-frequency signal comprising 440 and 480 kHz, leading to a noticeable beat. 
\item[(ii)] \textbf{Overtones.} Some frequency diagrams show distinct overtones and side lobes. Figures \ref{fig:spectrum_de2}, \ref{fig:spectrum_de1}, and \ref{fig:spectrum_ro3} present a few examples. 
\item[(iii)] \textbf{Duty cycle.} Except for one network, all European providers use four seconds off-time and one second on-time. Only AT-A1 has an off-time of five seconds in their circuit-switch service but four seconds off-time with VoLTE. In contrast, all tested North American networks used a two seconds on-time and four seconds off-time.
\item[(iv)] \textbf{Amplitude.} The amplitude can differ between networks by 10 dB (Figure \ref{fig:ringback-fingerprint}). 
\end{enumerate}
The identified features show high stability between experiments, with two exceptions. 
Some providers had two different amplitude levels (or frequency settings) and required us to make more test calls and recordings. This might be an artifact of multiple load-balanced switches being configured slightly differently.
Besides our North American samples, three out of 92 European test calls experienced a non-repeatable U.S.-style 440/480 Mhz ringback tone. We might have observed an attempted interconnect bypass fraud attack, as described by Sahin et al. \cite{sahin2016over}, or a case where the early media stream prematurely terminated at an earlier voice switch (e.g., transient connection errors).

For Figure \ref{fig:ringback-fingerprint}, we followed an AT\_A1 SIM card across 19 providers in nine countries -- plotting only frequencies and signal strengths. Dual frequencies are suffixed with an \_L and \_H for low and high tones, respectively. 
In the case of AT\_A1, all four markers change as soon as the SIM card leaves its home operator. 
A few cases of (near) collision exist between HR\_A1 \& SL\_A1 as well as FI\_Telia \& RO\_Digi. 
Clashes between operators with multiple frequencies (e.g., RO\_Orange \& SK\_Telekom) are easily distinguished with repeated calls: The alternative frequencies do not clash. In contrast, the North American operators only differ in amplitudes.

VoLTE roaming is currently in its infancy, i.e., non-functional with many operators. They either use CSFB to switch to legacy voice or inform their roaming customers that voice is not available (e.g., AT\&T in the U.S.). 
Domestic VoLTE calls showed slightly different signal compositions. 

\paragraph{Charges.}
Between operators and other wholesale, phone calls are only charged after they were successfully established (e.g., the callee answered) and then typically on a per-second basis. Most operators pass the former modality to their customers but round up call durations to whole minutes. For calls long enough to be redirected to voice mail, there is a passive charge (for the incoming call) and an active charge for redirecting the call to the voice mail at the home operator. 
However, we also found that at least one operator charged for roaming voice calls to and from AT\&T, even though AT\&T does not support voice roaming.

\section{Discussion and Future Work}\label{sec:discussion}

Net neutrality, traffic discrimination, and Internet censorship are ongoing research topics. For example, some providers use questionable techniques to throttle video streaming~\cite{li2019large}.
Although the scientific community tries to address these issues, roaming scenarios are notoriously hard to measure.

In our first use case we discovered two traffic discrimination cases in roaming scenarios. 
Under the EU ``roam like at home'' doctrine~\cite{eu-reg-2017-920}, this traffic should be treated equally.
In our second use case we extend the idea to hide traffic from the provider's accounting in \textit{included services}.
This opens possibilities to tunnel arbitrary traffic without metering.

Further research opportunities can be found in (i) measuring more roaming user discrimination, (ii) fingerprinting roaming artifacts to identify a roaming user's operator and country, and (iii) reverse-engineering zero-rating identification techniques employed by operators~\cite{gegenhuber2022zero} (we only tested SNI and host headers).

Inspecting proactive APDU traffic is highly relevant for the privacy and security: 
The SIM microcontroller is much more than just a configuration storage or a cryptographic co-processor. 
Proactive or otherwise hidden SMS and Internet connections are opaque to the user and run in the background. 
Their actions can have negative effects on the monthly bill. 

We can expect eSIMs to behave even more intransparently since they are provisioned and updated over the air. 
Their SIM card communication is not inspectible by current physical MITM tools \cite{simtrace}. 
\toolname's rSAP and remote SIM bridge closes this gap. 
Apart from proactive communication, we also experienced SIMs changing their IMSI on the fly for roaming purposes.
We expect even more IMSI provisioning from \textit{travel SIMs} with flexible country/region data plans. 

\toolname provides a robust and versatile testbed and measurement platform for such investigations. 
We tested our framework with round-trip times of up to 1,000 ms.
In comparison, geostationary Internet connections typically have round-trip times between 300 and 600 ms. 
Therefore we are confident the technique is robust enough to tunnel any SIM to any point on earth.

\paragraph{Expansion}
Our framework currently consists of twelve probes that cover eight European and two North American countries. %
We open sourced the design and specifications of our probe to enable other researchers to participate.
While budgetary constraints originally motivated the low-cost design, it now allows the project to scale quickly.
We continuously seek organizations to host probes and provide on-the-ground support with SIM acquisitions. \looseness=-1

\paragraph{User groups}
We identified multiple user groups -- in addition to researchers -- who might be interested but are currently not considered: (i) Internet activists, (ii) regulatory authorities, and (iii) ISPs.
Due to its open and low-cost design, Internet activists and watch groups could easily use \toolname to test networks themselves.
By hosting their own testing infrastructure based on our technique, network operators could also benefit, e.g., by testing characteristics and interplay with foreign partner networks.
Still, \toolname is currently adapted to research purposes and, in the first step, we plan to collaborate with other researchers.

\paragraph{Research directions} 
Future additional measurement and testbed uses can be, but are not limited to (i)~cellular infrastructure scans, (ii)~SIM vulnerability analyses, (iii)~IPv4/IPv6 deployment surveys, (iv)~Quality of Experience measurements, (v)~geoblocking configurations, (vi)~roaming traffic speed discrimination, and (vi)~OTT and PBX fraud detection.

\paragraph{Improve SIM tunneling}
Certain known %
APDU requests, e.g., static files, should be emulated locally.
This includes many local settings, such as the phone book or the list of last calls.
This will speed up initialization and connection times as it eliminates round-trips over the network.

\paragraph{Probe hardware iteration with 5G}
A 5G prototype is in development and will be deployed when funding is secured.

\paragraph{Mobile measurement probes} %
Currently, the probes are used in a stationary manner.
With a proper control channel (e.g., Wi-Fi) and sufficient power supply (e.g., in trains or buses), the probes could also be used in a mobile scenario to better reflect user behavior or map network properties geographically.

\subsection{Limitations}

\paragraph{Low-level radio control} \toolname lacks the radio versatility of an SDR. 
However, such a setup would incur radio-regulatory challenges and was therefore out of scope for this paper. 

\paragraph{Detectability} Certain behaviors (e.g., fast geographical movements of SIMs) could trigger fraud detection systems. So far, we did not experience that. 

\paragraph{Test automation}
The current implementation lacks the anticipated versatile multi-tenant scheduling system as test scripts have to be run semi-automated.

\paragraph{SIM card management}
The proposed technique vastly reduces hardware and operational complexity and costs.
However, managing and maintaining the SIM cards and phone plans remains a time consumer. 

The cards have to be purchased, registered, and regularly topped up, in connection with multiple different systems and in different languages.

Many operators do not ship their SIMs internationally, so we need to build a network of partners to either send SIM cards to us or host them locally.
So far, eSIMs are not universally offered and may only alleviate some of the efforts.

\section{Conclusion}\label{sec:conclusion}

Many current measurement frameworks and tools insufficiently capture (a) the unique properties of cellular networks and (b) their emergent complexities arising from interconnecting them internationally (with all the entailing security and privacy risks). 
\toolname overcomes access technology obliviousness, the quadratic scaling of resources, and background noise challenges. The improved scalability and automation allow for measurements and investigations deemed unpractical before. 

To this end, we geographically decoupled radio devices from the subscriber module by taking advantage of the low-level control in SIM communication.
Our TCP tunneling technique proved to be robust under high-latency conditions, allowing for worldwide use.
Furthermore, \toolname provides a high level of data hygiene by decoupling the measurement from any other IP traffic. 
For billing-related research, \toolname offers integrated support for various data accounting interfaces. The low-cost and open-source design makes it ideal for decentralized deployment. All source material and schematics will be available upon publication.

We presented the capabilities in security and privacy research through five exemplary use cases: 
(i) roaming and price differentiation, 
(ii) systematic traffic accounting experimentation with the ability to free-ride traffic,
(iii) large-scale network property analysis,
(iv) low-level APDU analytics, and
(v) fingerprinting call characteristics to pinpoint subscribers.

In all cases, we found surprising results. Despite the fact that all operators in our sample use home routing, differences in the metering of roaming connections occur. With those use cases, we demonstrated how to send unaccounted traffic, i.e., \textit{free-riding} or \textit{phreaking}. 

Furthermore, in many cases, minuscule differences in call progress tones leak the callee's visiting network operator -- enabling a country-level localization.
Additionally, some SIM cards proactively and covertly send binary SMS to their home operators; however, their analysis is inconclusive so far.

To lower the barrier for large-scale security and privacy measurements in cellular networks,
we plan to expand the network to more researchers and countries. Additionally, we open-sourced all software and schematics\footnote{\url{https://github.com/sbaresearch/mobile-atlas}}.

\section{Acknowledgments}
We would like to thank the reviewers and shepherd of this conference for their constructive feedback, but also those of NDSS, MobiCom and IMC.

This work were not possible without the hosters of our probes. 
We heartily thank 
Giovanni Camurati (EURECOM, now ETH Zurich),
Aurélien Francillon (EURECOM),
Michael Franz (UCI),
Katharina Krombholz (CISPA),
nakano,
Sebastian Sieh,
Tanja Šarčević (SBA Research),
Daniel Sokolov (Heise Online),
and six more unnamed probe hosters.

We also like to thank Harald Welte for an experience exchange early in the project.

Financing for cellphone research in Austria proven very challenging over the years. Therefore we are very pleased that the this project was funded through the NGI0 PET Fund, a fund established by NLnet with financial support from the European Commission's Next Generation Internet programme, under the aegis of DG Communications Networks, Content and Technology under grant agreement No. 825310.
Furthermore the awards from Office of Naval Research, N00014-22-1-2232 and N00014-21-1-2409, DARPA I2O Small Business Technology Transfer (STTR) Program, W31P4Q-20-C-0052, Austrian Science Fund (FWF) P30637-N31 partly supported this work.
The competence center SBA Research (SBA-K1) is funded within the framework of COMET--Competence Centers for Excellent Technologies by BMVIT, BMDW, and the federal state of Vienna, managed by the FFG.

\let\oldbibliography\thebibliography
\renewcommand{\thebibliography}[1]{%
  \oldbibliography{#1}%
  \setlength{\itemsep}{0pt}%
  \setlength{\parskip}{1pt plus 2pt}
}

\bibliographystyle{plain}
\bibliography{mobileatlas,adrianmobile,rfc}

\appendix

\section{Probe}
\label{subsec:probe}
\subsection{Hardware}

\begin{figure}[!t]%
    \centering%
    \includegraphics[width=.90\linewidth,trim=20px 20px 20px 20px, clip=true]{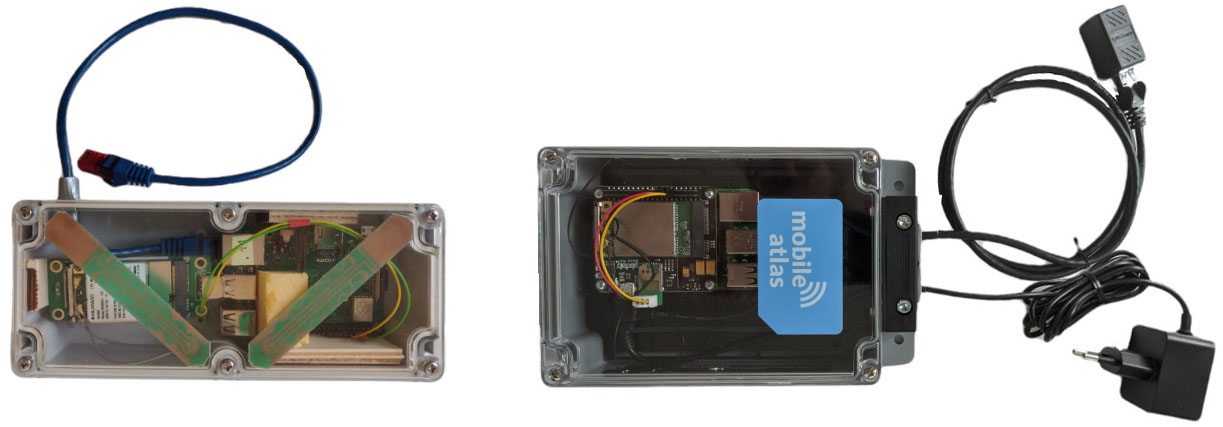}%
    \caption{Left: first prototype; right: current version.}%
    \label{fig:probe-version-comparison}%
    \vspace{-3mm}
\end{figure}

\toolname probes execute the measurements.
Each probe is deployed on a single-board computer that provides an Internet uplink, a USB host support, and a general-purpose in- and output (GPIO) interface that is used for SIM card emulation.
The single-board computer is connected to a modem that provides access to the cellular network.

We chose the \emph{Raspberry Pi 4} as a cost-friendly platform with more than sufficient computing power for our purpose.
After comparing various modems, we selected the \textit{Quectel EG25-G} modem; it has good ModemManager support and tolerates the imprecise timing of our universal asynchronous receiver-transmitter (UART).
Since modern modems are designed for notebooks and embedded systems, we used an additional \textit{USB -- Mini PCIe adapter}.
In order to facilitate user-friendly deployment, we packaged the probe with all modules and antennas into a single case with just two connectors, namely for RJ45 Ethernet and power.
The selection for this study was driven by cost-efficiency, and our software is optimized for -- but not restricted to -- the components described above.

We developed an initial prototype and refined our design in a second version, as presented in Figure~\ref{fig:probe-version-comparison}.
We found and solved several issues by testing our prototype in multiple field tests.
First, we improved the modem adapter and modem mounting and changed the casing, as there were problems due to rough handling during  shipping.
Second, we introduced a SIM adapter printed-circuit board \ifdefined\BLINDED\else(see Figure~\ref{fig:sim-adapter})\fi, instead of soldering directly on the SIM slot.
Further, we updated the \emph{Raspberry Pi 3} (used for the prototype) to the current version 4.
Finally, we exchanged the Huawei ME909 modem with a Quectel EG25-G for worldwide radio band coverage and better {ModemManager} support thanks to its popularity in the open-source community\footnote{The EG25-G modem is also used in the open-sourced PinePhone.}.

\subsection{Software}
\begin{figure}[!t]%
    \centering
    \includegraphics[width=.90\linewidth]{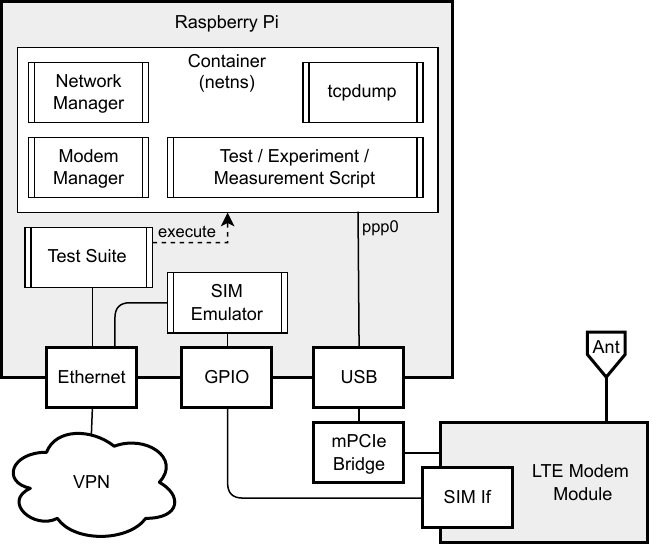}
    \caption{Probe architecture: GPIO ports physically emulate the SIM card to the modem. Separate network namespaces eliminate unwanted background traffic (e.g., from the OS or other system services) in experiments.}
    \label{fig:probe-architecture}
    \vspace{-2mm}
\end{figure}

An overview of the probe's software architecture is presented in Figure~\ref{fig:probe-architecture}.
All components were implemented in Python~3 for rapid prototyping.
The low-level protocol communication with the SIM card and modem was built on pySim~\cite{pysim-project}.
For a specific measurement or experiment, the \emph{probe} requests a particular SIM card from the management system (using its IMSI) and sets up a TCP connection to the appropriate \textit{SIM provider}.
The SIM provider accepts it and bi-directionally forwards \textit{application protocol data unit} (APDU) traffic.

On the probe's side, a Linux network namespace binds only selected processes to the cellular network connection.
Within the namespace, {ModemManager} and {NetworkManager}~\cite{network-manager} handle the initialization of the modem connection and the networking environment.
Next, the measurement or experiment functionality is executed within the namespace and its separated network stack.
Any direct test case results, as well as the logs of ModemManager and NetworkManager, are stored in a JSON file, and a tcpdump instance records the occurring traffic for (optional) later analysis.
Additionally, tunneled APDU traffic is stored in a separate database and can be exported as a \emph{GSM SIM} PCAP file for later analysis.

In order to automate the credit checking for tests that involve billing, a feasible way to query the current quota needs to be manually implemented for each provider or tariff.
The \toolname software provides predefined checking procedures and a rich set of methods (e.g., network, SMS, USSD) that can be utilized to access this information in an automated and iterative manner.
This reduces manual intervention during test execution.

\subsection{SIM Interface}\label{sec:sim-interface}

After power-on, the SIM waits for a rising edge on the reset (RST) pin.
Since the specification allows different SIM operating voltages (1.8V, 3V, and 5V), the modem increases the voltage and resends the RST signal if no response was received.
When an appropriate RST signal is detected, the SIM reacts with an \textit{Answer To Reset} (ATR) message containing various card characteristics (e.g., maximum supported clock frequency).
After that, the modem can (but does not have to) perform a \textit{Protocol and Parameter Selection} (PPS) command to change the current connection properties.
Our modems use the fastest offered rate; without a PPS they stay at the initial speed. \looseness=-1

Data (i.e., the APDUs) are sent via a bi-directional open-collector I/O line with the modem-provided clock (CLK) signal divided by the negotiated divider (ATR or PPS).

\begin{figure}[!t]%
    \centering
	\includegraphics[width=1\linewidth]{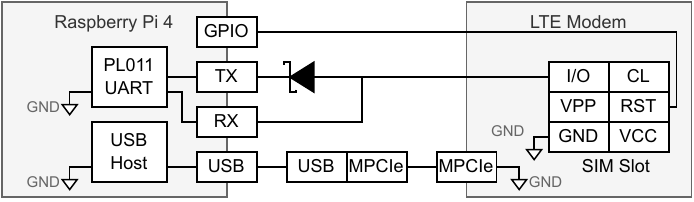}
    \caption{Wiring diagram between the SoC and the modem. The UART emulates SIM cards placed at remote locations.}
    \label{fig:sim-interface}
    \vspace{-3mm}
\end{figure}

Because the UART of our single board computer does not offer a synchronous mode (i.e., a USART), we approximate it through an oscilloscope measurement and program the UART's baud rate accordingly.
We emulate an open-collector output with the help of a Schottky diode (Figure~\ref{fig:sim-interface}).
The modem already provides a necessary pull-up resistor.

Generous divider settings, tolerances, and the ability to start communication at any clock cycle work in our favor and reduce cost.
Additionally, our SIM emulator can repeatedly signal \textit{Waiting Time eXtensions} (WTX) to cover for longer network round-trip times.

An additional GPIO port serves the SIM's RST pin.
For the sake of simplicity, we left out the ground wire and reused the USB's ground path, leading to slightly increased noise levels.

As our SIM tunnel implementation is aware of the ISO7816 protocol, we can independently choose connection parameters via ATR and PPS on both sides of the tunnel.
We can trigger a PPS procedure %
from the modem when we present an ATR with relatively fast settings. %
Alternatively, we can signal the same settings that were used for the initial ATR transmission
 so the modem skips PPS.

\section{SIM Provider}
\label{subsec:sim-provider}

\ifdefined\BLINDED
\else
\begin{SCfigure}[2][!b]%
    \centering%
	{\includegraphics[width=.28\linewidth]{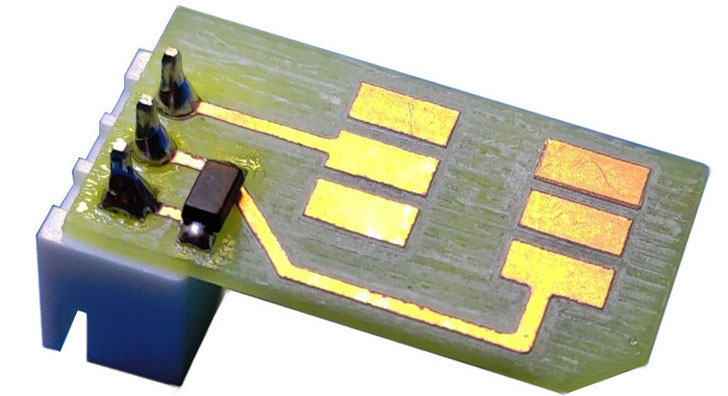}}
    \caption{SIM adapter is plugged into the modem's socket.}
    \label{fig:sim-adapter}
\end{SCfigure}
\fi

The TCP tunnel between a SIM emulator process on the probe and the SIM provider forms a virtual circuit in which one SIM can be connected to precisely one modem at a time.

\emph{SIM providers} are responsible for connecting the actual SIM cards to the measurement framework.
There exists dedicated hardware to connect a multitude of different SIM cards. The so-called SIM banks are sometimes used for OTT VoIP gateways.
These devices are often expensive, proprietary, and do not perfectly reflect our use case.
Instead, we chose the cost-efficient approach of connecting SIM cards with individual SIM readers.
These devices are cheap, readily available, and connect via USB.
Our system supports cheaper USB-serial-type devices as well as full-fledged PC/SC readers.

Although the USB standard theoretically allows for up to 127 connected devices, we experienced non-deterministic failures on multiple hosts (enumeration failure, USB driver errors) when chaining multiple USB hubs and connecting an \textit{unusually} high amount of devices to the USB bus.
We tested multiple hardware configurations, and since some devices (e.g., Raspberry Pi, Intel NUC, cheap USB hubs) did not work reliably, we ended up using a commodity laptop computer (in combination with multiple FE 2.1. hubs). %
We were able to successfully test our SIM provider for up to 32 concurrently attached SIM cards via one USB host.
The framework allows us to add any additional SIM cards from other distributed sources, although we currently obtain all SIM cards from one SIM provider.
With this distributed approach, we facilitate the testing of a multitude of mobile contracts, as one could temporarily add a SIM card and test it on a large scale.

\subsection{Bluetooth SIM Access Profile}\label{subsubsec:impl-bluetoth-sim-access}
As an alternative to USB-based readers, we also support sharing a SIM card via Bluetooth's (remote) SIM Access Profile (rSAP).
This legacy protocol was used for automotive integration, allowing external devices without a SIM card (such as cars) to access a phone's SIM card.

Our SIM provider can act as an
SAP client to a smartphone's rSAP server. %
On Android, new rSAP connections require user approval and only serve the primary SIM card, i.e., a possible eSIM needs to be configured as the primary SIM card.
Our tests run successfully on an off-the-shelf, non-rooted Google Pixel~3a. %

\section{Benchmarking}
\label{subsec:benchmarking}

In this Section, we evaluate the compatibility of different modem- and SIM reader models.

\subsection{Different modems}\label{subsubsec:different-modem-types}

We tested five different modem types (four state-of-the-art LTE modems, one legacy 3G device) and compared their characteristics, as shown in Table~\ref{tab:modem-functionality}.
Since we want our framework to utilize the full spectrum of mobile phone functions, we tested key features of every modem via the ModemManager interface.
\toolname supports network connections, calling, SMS, and sending AT commands with all of the tested modems.
USSD codes work with four tested modems, whereby two modems required a small patch to set the correct character encoding during the setup routine.
The remaining modem accepts USSD directly via AT commands, but not via the ModemManager interface.
Furthermore, we benchmarked the time needed for setting up a successful mobile network connection with \toolname in a domestic setting with good reception and a low-latency connection.
Although the average time differs, it stays below one minute for all models.%

Table~\ref{tab:modem-sim-tunnel} summarizes the SIM tunneling characteristics for these five modems.
All modems but one support tunneling with a faster PPS-negotiated baud rate.
The initialization procedure and the number of requested APDUs differ vastly between modems.
Telit's modem (M4) is a clear outlier because it requests large parts of the SIM's file system at startup.

To determine the right baud rate for our SIM tunnel interface, we measured each modem's clock with an oscilloscope.
Furthermore, we checked how tolerant the modems are to APDUs that were based on deviant clock speeds.
Our measurements show that the tested modems are fairly tolerant and the SIM tunnel's base clock can diverge by around 4\% without having any negative impact on the APDU communication.

\begin{table}[!t]%
\newcommand{\modemh}[1]{\rot{#1}}%
\caption{Supported functionality \\ for different modems \\ (via \texttt{option} driver).}%
\label{tab:modem-functionality}%
\footnotesize%
\vspace{-50pt}
\begin{tabular}{|p{3cm}|X{.5cm}X{.5cm}X{.5cm}X{.5cm}X{.5cm}|}
\multicolumn{1}{l}{}	 & \modemh{Quectel EG25-G} & \modemh{Huawei ME909s-120} & \modemh{SIMCom SIM7600E-H} & \modemh{Telit LE910} & \modemh{Sierra MC8355} \\
\cline{2-6}
\multicolumn{1}{l|}{}    & M1        & M2         & M3         & M4          & M5         \\
\hline                        
Network                  & \checkmark & \checkmark & \checkmark & \checkmark & \checkmark \\
Connection time (sec)    & 28         & 58         & 29         & 51         & 51         \\
Start/Stop Call          & \checkmark & \checkmark & \checkmark & \checkmark & \checkmark \\
Send/Receive SMS         & \checkmark & \checkmark & \checkmark & \checkmark & \checkmark \\
Send/Receive USSD        & $\sim$     & \checkmark & $\sim$     & $\times$   & \checkmark \\
Send AT Cmd              & \checkmark & \checkmark & \checkmark & \checkmark & \checkmark \\
    \hline
\end{tabular}
\begin{minipage}[l]{0.8\linewidth}\scriptsize
\vspace{1mm}
$\sim$ works with adjusted encoding
\vspace{1mm}
\end{minipage}
\end{table}

\begin{table}[!t]
\vspace{-3mm}
\caption{Our SIM tunneling works well across all tested COTS modem models.}\vspace{-1ex}%
\label{tab:modem-sim-tunnel}%
\vspace{-2mm}%
\footnotesize%
\begin{tabular}{|p{3cm}|X{.5cm}X{.5cm}X{.5cm}X{.5cm}X{.5cm}|}
\cline{2-6}
\multicolumn{1}{l|}{}      & M1        & M2         & M3         & M4          & M5         \\
\hline
Tunnel without PPS        & \checkmark    & \checkmark   & \checkmark   & \checkmark  & \checkmark   \\
Tunnel with PPS           & \checkmark    & \checkmark   & \checkmark   & $\times$         & \checkmark   \\
Initialization (sec)     & 24            & 49           & 25           & 131         & 51           \\
Initialization (\#APDUs) & 261           & 638          & 271          & 1913        & 501          \\
SIM clock (MHz)           & 3.842         & 3.759        & 3.842        & 3.250       &  3.842       \\
Clock tolerance (neg \%)  & 4.1           &   3.8        & 4.1          & 4.2\textsuperscript{\textdagger} & 4.1 \\
Clock tolerance (pos \%)  & 4.5           &   4.9        & 4.5          & 5.2\textsuperscript{\textdagger} & 4.4 \\
\hline
\end{tabular}
\begin{minipage}[l]{0.8\linewidth}\scriptsize
\vspace{1mm}
\textsuperscript{\textdagger}~without PPS
\vspace{1mm}
\end{minipage}
\end{table}

\subsection{SIM readers}\label{subsubsec:sim-provider-analysis}

SIM providers rely on pySim~\cite{pysim-project} in combination with hardware SIM readers, with either PC/SC (CCID) interfaces or a simple USB serial interface.
The former are more expensive (USD 10--30), but are typically more versatile.
Several low-cost readers (USD 2--5) provide a simplistic serial interface via USB (\emph{USB Serial Adapter}).
We tested seven SIM readers with 14 different SIM cards from five countries, all purchased in 2021.
The same SIM cards are deployed in our use cases in Section~\ref{subsec:roaming-and-price-differentiation}.
As Table~\ref{tab:sim-reader-compatibility} shows, three of the SIM readers worked with every SIM card.
Two SIM readers had problems with one specific SIM card.
One particular low-cost SIM reader had problems with several SIM cards in PPS mode, but accepted 12 out of 14 SIMs without PPS.

In our sample, PC/SC devices processed APDUs faster than the low-cost serial-based SIM readers.

\begin{table}[!b]%
\vspace{-5mm}
\caption{SIM reader compatibility}
\centering
\label{tab:sim-reader-compatibility}
\footnotesize
\begin{tabular}{|lr|}
\hline
Modem                               & Supported SIMs \\
\hline
Gemalto HWP108841HY                 & 14 / 14 \\
Manhattan 102049                    & 14 / 14 \\
Woxter CN43296012                   & 14 / 14 \\
Bit4id MiniLector EVO               & 13 / 14 \\
Low-cost SIM reader, White          & 13 / 14 \\
Low-cost SIM reader, Blue           & 7 / 14  \\
Low-cost SIM reader, Blue (w/o PPS) & 12 / 14 \\
\hline
\end{tabular}
\end{table}

\end{document}